\documentclass[aps,prl,preprint,groupedaddress]{revtex4}
\usepackage{epsfig}
\usepackage{bm}

\newcommand{\be}{\begin{equation}}
\newcommand{\ee}{\end{equation}}
\newcommand{\ba}{\begin{eqnarray}}
\newcommand{\ea}{\end{eqnarray}}

\newcommand{\ds}{\displaystyle}
\newcommand{\scs}{\scriptscriptstyle}

\newcommand{\qvec}{\mbox{\boldmath $q$}}
\newcommand{\kvec}{\mbox{\boldmath $k$}}
\newcommand{\Kcapvec}{\mbox{\boldmath $K$}}

\newcommand{\pvec}{\mbox{\boldmath $p$}}
\newcommand{\Pcapvec}{\mbox{\boldmath $P$}}

\newcommand{\Svec}{\mbox{\boldmath $S$}}

\newcommand{\sigmavec}{\mbox{\boldmath $\sigma$}}

\begin{document}

\title{$\Lambda(1520)$  and $\Sigma(1385)$ in the nuclear medium}

\author{Murat M. Kaskulov} 
\email{kaskulov@ific.uv.es}
\author{ E. Oset}
\email{oset@ific.uv.es}
\affiliation{Departamento de F\'{\i}sica Te\'orica and IFIC,
Centro Mixto Universidad de Valencia-CSIC,
Institutos de
Investigaci\'on de Paterna, Aptd. 22085, 46071 Valencia, Spain}
\date{\today}

\begin{abstract} 
Recent studies of the $\Lambda(1520)$ resonance within chiral unitary theory
with coupled channels find the resonance as a dynamically generated state from
the interaction of the decuplet of baryons and the octet of mesons, 
essentially a quasibound state of $\pi \Sigma^*(1385)$ in this case, 
although the coupling of the $\Lambda(1520)$ to the $\bar{K}N$ 
and $\pi \Sigma$ makes this picture only approximate. 
The $\pi \Sigma^*(1385)$ decay channel of
the  $\Lambda(1520)$ is forbidden in free space for the nominal mass of the
$\Sigma^*(1385)$, but the coupling of the $\pi$ to $ph$ components 
in the nuclear medium  opens new decay channels of the $\Lambda(1520)$ 
in the nucleus and produces a much larger width. Together with medium 
modifications of the $\bar{K}N$ and $\pi \Sigma$ decay channels, 
the final width of the $\Lambda(1520)$ at nuclear matter density is 
more than five times bigger than the free one. We perform the
calculations by dressing simultaneously the $\Lambda(1520)$ and the
$\Sigma^*(1385)$ resonances, finding moderate changes in the mass but 
substantial ones in the width of both resonances.
\end{abstract}
\maketitle 

\section{Introduction}
\small

The $\Lambda(1520)$ is an intriguing resonance which has captured much 
attention and is easily produced in $K^-$ induced reactions 
\cite{Chan:1972nj,Mast:1972wc,Mast:1973cz,Berley:1996zh} or photon induced
reactions \cite{Barber:1980zv}.  Two recent different initiatives have 
brought this resonance to be again a focus of attention.  On the one hand, 
this resonance appears, and is an important reference, in  experiments that 
try to see  the pentaquark $\Theta^+$ \cite{nakano} (see \cite{hyodo} for 
a detailed reference of papers on this issue and \cite{hicks} for a recent 
review). In fact, the large background
appearing in the tail of the $\Lambda(1520)$ at energies higher than the 
nominal mass \cite{pentatalk} is one of the issues to be clarified when 
trying to make analyses on the $\Theta^+$. The issue of the large background
has already been addressed in \cite{luis} and it is found to be associated 
to the large coupling of the resonance to the $\pi \Sigma^*(1385)$ channel.  

The other initiative concerning the $\Lambda(1520)$ has been the study of 
the interaction of the decuplet of baryons with the octet of pseudoscalar 
mesons \cite{kolo,sarkar}, which has brought as an output that many of the 
low lying $3/2^-$ resonances are dynamically generated from the interaction 
of the coupled channels of these two multiplets. In particular, 
the  $\Lambda(1520)$ appears basically as a quasibound state of the 
$\pi \Sigma^*(1385)$ system. The small free width of the 
$\Lambda(1520)$ of $15$~MeV comes from the decay into  $\bar{K}N$
and $\pi \Sigma$, since the decay into $\pi \Sigma^*(1385)$ is forbidden 
for the
nominal mass of the $\Sigma^*(1385)$.   Of course, the coupling of the 
$\Lambda(1520)$ to $\bar{K}N$ and $\pi \Sigma$ 
makes the picture of the $\Lambda(1520)$ more elaborate, with 
$\pi \Sigma^*(1385)$ being a very important component but with also
sizable admixtures of $\bar{K}N$ and $\pi \Sigma$~\cite{souravreac,RSO}

  The change of resonance properties in the nuclear medium is also a field that
captures permanent attention, and basic symmetries can be tested through medium
modification of particle properties
\cite{osetweise,mosel,friman,chanfray,wambach,danielrho,angelsk,
danielphi,rapp}. The decay of the $\Lambda(1520)$ in the nuclear medium 
bears resemblance to the one of the $\Delta(1232)$ \cite{loren}. 
The  $\Delta$ decays into
$\pi N$ and the $\pi$ gets renormalized in the medium by exciting $ph$ and
$\Delta h$ components, as a consequence of which the $\Delta$ is renormalized
and its pion (photon) induced excitation in nuclei incorporates now the
mechanisms of pion (photon)
absorption in the medium. In the present case, the $\Lambda(1520)$ decay into
$\pi \Sigma^*(1385)$, only allowed through the $\Sigma^*(1385)$ width, gets
drastically
modified when the $\pi$ is allowed to excite $ph$ and $\Delta h$ components in
the nucleus, since automatically the phase space for the decay into $ph
\Sigma^*(1385)$ gets tremendously increased. This fact, together with the
large coupling of the $\Lambda(1520)$ to the $\pi \Sigma^*(1385)$ channel
predicted by the chiral theory, leads to a very large width of the
$\Lambda(1520)$ in nuclei. Similar nuclear effects will
modify the $\pi \Sigma$ decay channel and the $\bar{K}N$ will be analogously
modified when $\bar{K}$ is allowed to excite hyperon-hole excitations. All
these channels lead to a considerable increase of the width of the 
$\Lambda(1520)$ in the nucleus.

Concerning the medium corrections of the $\pi \Sigma^*(1385)$ channel, this
has already a precedent in physics in the decay of an ordinary
$\Lambda(1115)$ in nuclei. The free $\Lambda(1115)$ decay into $\pi N$
through weak interactions, mesonic decay, is largely suppressed by
Pauli blocking in nuclei. However, the pion can excite $ph$ components in the 
medium
leading to a new $\Lambda(1115)$ decay mode $\Lambda(1115) \to N ph$, 
or equivalently
$\Lambda(1115) N \to NN$, non mesonic decay. 
This new decay channel is far bigger
than the mesonic decay in the nucleus and as large as the free 
one~\cite{Millener:1988hp,Motoba:1988sk,Oset:1989ey,Alberico:2001jb,
Grace:1985yi,Sakaguchi:1990wd}.

The
confirmation of the large width of the $\Lambda(1520)$ in the medium, 
more than a factor five times
larger than the free one as predicted here, would provide a strong support for
the nature of the  $\Lambda(1520)$ as a dynamically generated resonance 
from $\pi \Sigma^*$, $\pi \Sigma$ and $\bar{K}N$ channels. Clear indications
that this might be the case can be seen in the analysis of $\Lambda(1520)$
production in heavy ion reactions~\cite{R1,R2}.

We have organized the paper as follows. In Sections II and III 
the model for the 
$\Lambda(1520)$ and $\Sigma^*(1385)$ self energies in the nuclear medium 
is described. The result and discussions are presented in Section IV and V.
Finally, the conclusions are given in Section VI.

\section{Renormalization of the $\Lambda(1520)$}

In this section we discuss the formalism used in the present work
for the description of in-medium properties of the $\Lambda(1520)$ hyperon.
Here we follow the standard approach  where the nuclear medium  is 
described 
by the  noninteracting Fermi sea
and the baryonic resonances get modified in the nuclear medium in the 
dressing procedure
by coupling the mesons in the loops to baryon - and hyperon - hole 
excitations.    


\subsection{The $S$-wave decay of the $\Lambda(1520) \to \pi \Sigma^*(1385)$}

In the study of ~\cite{kolo,luis,souravreac} the $\Lambda(1520)$ is generated
dynamically from the $\pi \Sigma^*(1385)$ and 
$K \Xi^*$ channels interacting in 
$S$-wave. In~\cite{souravreac} the $\bar{K}N$ and $\pi \Sigma(1189)$ 
channels in
$D$-wave are added in order to produce the proper width of the
$\Lambda(1520)$. Of all these channels the most important one is the 
$\pi \Sigma^*(1385)$, but the width into this channel is very 
small due to largely
reduced phase space for the $\Lambda(1520)$ decay into $\pi\Sigma^*(1385)$, 
only
possible through the tail of the $\Sigma^*(1385)$ when 
its width is 
considered. 
But the 
relevance of the $\pi \Sigma^*(1385)$ channel and the fact that in the nuclear
medium the phase space for decay into this and associated channels becomes very
large, makes the renormalization of this channel very 
important in studying the $\Lambda(1520)$ in the nuclear medium.

\begin{figure}[t]
\begin{center}
\includegraphics[clip=true,width=0.65\columnwidth,angle=0.]
{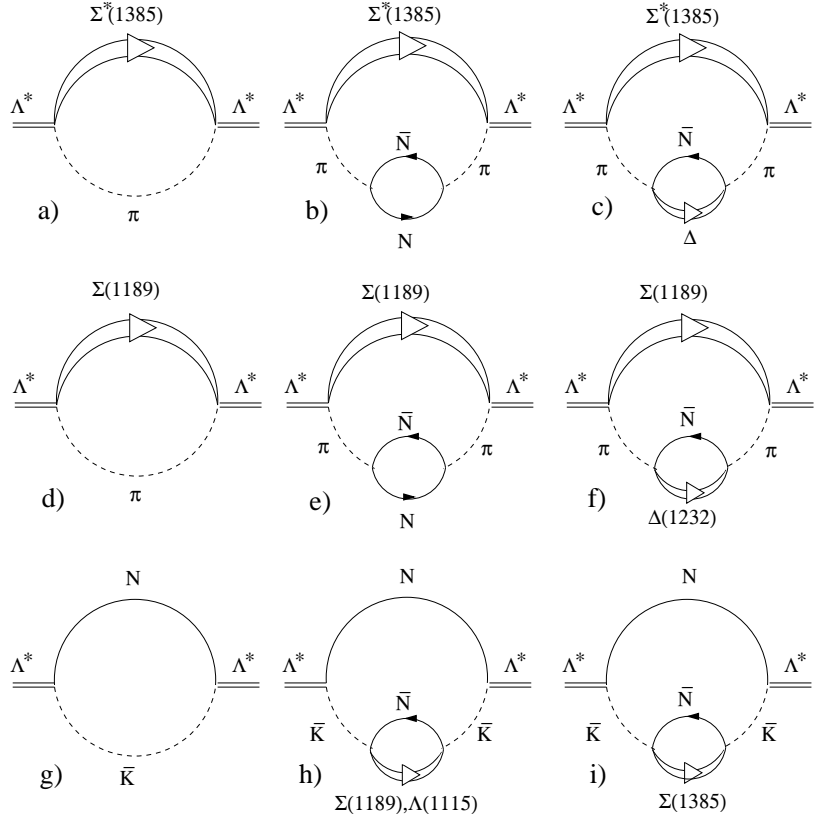}
\caption{\label{Lambda1} \footnotesize The renormalization of the 
$\Lambda(1520)$ in the nuclear medium. }
\end{center}
\end{figure}

The in-medium 
renormalization of $\Lambda(1520)$ in the $\pi\Sigma^{*}(1385)$
channel
can be represented by the first three diagrams 
in Fig.~\ref{Lambda1}. 
In these diagrams the $\Sigma^{*}(1385)$ arises
as an intermediate state but will be also dressed
in its relevant decay channels. 
Hence, in the present work we address simultaneously
the dressing of the $\Sigma^{*}(1385)$ and $\Lambda(1520)$ in the nuclear 
matter.

In the following we
specify the in-medium propagators of the hyperons $\tilde{D}_Y$ 
and pions $\tilde{D}_{\pi}$. 
The kaon propagation in the nuclear medium will be addressed
separately. 
In terms of a dispersion relation 
representation 
\begin{eqnarray}
\label{DY}
\tilde{D}_{Y}(K,\rho) &=& 
\int \limits_{0}^{\infty} d \, W
\frac{S_Y(W,\Kcapvec,\rho)}{K^0-W+i0^+}, \\
\tilde{D}_{\pi}(k,\rho) &=& 
\int \limits_{0}^{\infty} d\, \omega (2\omega) 
\frac{S_{\pi}(\omega,\kvec,\rho)}{(k^{0})^2-\omega^2 + i0^+}
\end{eqnarray}
Here $K(k)$ are the four momenta and $S_{Y(\pi)}$ are the 
spectral functions of hyperons (pions),
$\rho$ is the nuclear matter density
 and Eq.~(\ref{DY}) 
accounts for the
positive energy part of the fermionic propagator only. 
For $S_{Y(\pi)}$ we have
\begin{equation}
\label{SF}
S_{Y(\pi)} = - \frac{1}{\pi} \mbox{Im} \Big[ \tilde{D}_{Y(\pi)}\Big]
\end{equation}
where
\begin{widetext}
\begin{eqnarray}
\label{SigmaSE}
\mbox{Im} \Big[ \tilde{D}_{Y}(W,\Kcapvec,\rho) \Big] = 
\frac{M_Y}{E_{Y}(\Kcapvec)} 
\frac{\mbox{Im} \Big[ \Sigma_{Y}(W,\Kcapvec,\rho)\Big]}
{[W-E_{Y}(\Kcapvec)-\mbox{Re} \Sigma_{Y}(W,\Kcapvec,\rho)]^2
+[\mbox{Im} \Sigma_{Y}(W,\Kcapvec,\rho)]^2} 
\end{eqnarray}
\begin{eqnarray}
\label{PiSE}
\mbox{Im} \Big[ \tilde{D}_{\pi}(\omega,\kvec,\rho) \Big] = 
\frac{\mbox{Im} \Big[ \Pi_{\pi}(\omega,\kvec,\rho)\Big]}
{[\omega^2-\tilde{\omega}^2(\kvec)-\mbox{Re} \Pi_{\pi}(\omega,\kvec,\rho) ]^2
+[\mbox{Im} \Pi_{\pi}(\omega,\kvec,\rho)]^2}
\end{eqnarray}
In Eqs.~(\ref{SigmaSE}) and~(\ref{PiSE}) 
$E_{Y}(\Kcapvec)=\sqrt{\Kcapvec^2+M_Y^2}$ and 
$\tilde{\omega}(\kvec)=\sqrt{\kvec^2+m_{\pi}^2}$ are the on-mass-shell
energies
of hyperons and pions, respectively, and 
the in-medium self energy
$\Sigma_Y$ is the subject of  the
present calculations. The  $P$-wave pion polarization operator $ \Pi$ 
is given by 
\begin{eqnarray}
\label{PolOperator}
\Pi_{\pi}(k,\rho) &=& 
\left(\frac{D+F}{2 f_{\pi}} \right)^2 \kvec^2 
\mathcal{U}(k,\rho) 
\left[ 1 -  
\left(\frac{D+F}{2 f_{\pi}}\right)^2 g' \mathcal{U}(k,\rho) \right]^{-1}
\end{eqnarray}
\end{widetext}
where $D$ and $F$ are the 
axial vector coupling constants 
and $f_{\pi}=93$~MeV is the pion decay constant. One finds $F\simeq 0.51$, $D\simeq0.76$
and the axial coupling constant used in present calculations is 
$g_A=D+F \simeq 1.27$.
Also in Eq.~(\ref{PiSE}) $g'=0.7$ is the Landau-Migdal 
parameter~\cite{Migdal:1978az} and
$\mathcal{U}(k,\rho)=\mathcal{U}^d(k,\rho) + \mathcal{U}^c(k,\rho)$
is the Lindhard function  including the direct 
and crossed 
contributions of $p-h$ and $\Delta-h$ excitations 
with the normalization of the appendix of Ref.~\cite{Oset:1989ey}.
These are the conventional definitions. Latter on we
shall modify the formalism to account for the short-range (SR) 
correlations relevant for the in-medium dressing of the $\Sigma^*(1385)$.

The $S$-wave character of the
$\Lambda(1520)\to \pi \Sigma^*(1385)$ decay requires
the following transition amplitude
\begin{equation}
\label{TL}
-it_{\Lambda(1520) \to \pi \Sigma^*} = -i 
g_{\scriptscriptstyle \Lambda^{*}\pi \Sigma^*}
\end{equation}
where $g_{\scriptscriptstyle \Lambda^*\Sigma^*\pi}$ 
already accounts for the
three charge states in the isospin $I=0$ channel, see Eq.~(\ref{LSD}), 
in the decay $\Lambda(1520)\to \pi \Sigma^*(1385)$. 
Alternatively, one can use the isospin decomposition
(the convention for phases is from Ref.~\cite{souravreac})
\begin{eqnarray}
\label{LSD}
\Lambda(1520) \to 
&& 
|  \pi \Sigma^*(1385); I=0 \rangle  \nonumber \\
&& = \frac{1}{\sqrt{3}}
\Big[ 
| \pi^- \Sigma^{*+} \rangle -
| \pi^+ \Sigma^{*-} \rangle -
| \pi^0 \Sigma^{*0} \rangle
\Big] 
\end{eqnarray}
and multiply the coupling constant  
$g_{\scriptscriptstyle \Lambda^{*} \Sigma^* \pi}$ by 
Clebsch-Gordan coefficients $\pm \sqrt{1/3}$ in the corresponding 
vertices. The sum of partial decay width into these channels equals to the 
$I=0$ contribution provided by the single coupling of Eq.~(\ref{TL}).

The value of the coupling constant 
$g_{\scriptscriptstyle \Lambda^*\Sigma^*\pi}=1.57$ 
was determined in Ref.~\cite{souravreac} by means of the
residue of the scattering amplitude near the pole position 
of the $\Lambda(1520)$.  A value of 
$g_{\scriptscriptstyle \Lambda^*\Sigma^*\pi}=1.21$
was used in Ref.~\cite{luis} where the $\bar{K}N$ and $\pi\Sigma$ channels are
not considered. Recent studies which combine simultaneous description
of both the photoproduction and $\bar{K}N$ scattering data results in 
$g_{\scriptscriptstyle \Lambda^*\Sigma^*\pi}=0.89$~\cite{RSO}. The latter value is used
in the present work. Actually in Ref.~\cite{RSO} there is a discussion of
many reactions involving the $\Lambda(1520)$. However, for the purpose of
setting the strength of $g_{\scriptscriptstyle \Lambda^*\Sigma^*\pi}$
it suffices to quote that this coupling is needed to interpret the $K^-p
\to \pi^+ \pi^- \Lambda$ reaction \cite{Mast:1973gb} using the formalism
of Ref.~\cite{souravreac} and the empirical width of 15~MeV for the
$\Lambda(1520)$.
(The $K^-p \to \pi^+ \pi^- \Lambda$ cross section is twice as big as
that of $K^-p \to \pi^0 \pi^0 \Lambda$ studied in~\cite{souravreac}).

Using Eq.~(\ref{TL}) the one loop contribution to the $\Lambda(1520)$ 
selfenergy 
from $\pi \Sigma^*$ intermediate state takes the form
\begin{eqnarray}
\label{SELambda}
-i\Big[\Sigma(P,\rho)\Big]^{\Lambda(1520)}_{\pi {\it\Sigma}^*(1385)} 
&=& g^2_{\scriptscriptstyle \Lambda^*\Sigma^*\pi}  
\int \frac{d^4 k}{(2 \pi)^4} 
\tilde{D}_{{\it \Sigma}^*}(P-k,\rho) \tilde{D}_{\pi}(k,\rho) \nonumber
\end{eqnarray}
where $\tilde{D}_{\it \Sigma^*}$ and $\tilde{D}_{\pi}$ are 
in-medium propagators of the $\Sigma^*(1385)$ and pion, respectively.
The in-medium renormalization of the $\Sigma^*(1385)$ 
which enters Eq.~(\ref{SELambda}) 
will be addressed in Section~\ref{S2}.

\subsection
{The $D$-wave decay of the $\Lambda^*(1520) \to \pi \Sigma + \bar{K}N$}

The conventional $D$-wave decay of the $\Lambda(1520)$ into
the $\bar{K}N$ and $\pi \Sigma$ channels accounts for practically all of
the $\Lambda(1520)$ free width~\cite{Groom:2000in}. 
The diagrams responsible for the renormalization of the $\Lambda(1520)$ 
in these decay channels are shown in Fig.~\ref{Lambda1} (d-i).
The pertinent $D$-wave transition operator is given by
\begin{equation}
\label{VD}
-i t_{\pi \Sigma \to \Lambda^*(\bar{K}N \to \Lambda^{*})} = - i 
g_{\scriptscriptstyle \Lambda^* \pi \Sigma(\Lambda^* \bar{K} N)} 
(\Svec^{\dagger} \cdot \kvec) (\sigmavec \cdot \kvec)
\end{equation}
where $\kvec$ is the pion three momentum and 
$S_i^{\dagger}$ is the $2 \times 4$ transition operator from spin $1/2$ to spin
$3/2$ fulfilling the relation
$S_i S_j^{\dagger} = 2\delta_{ij}/3-i\epsilon_{ijk}\sigma_k/3$.
Both coupling constants
$g_{\scs \Lambda^* \pi \Sigma}$ and $g_{\scs \Lambda^* \bar{K} N}$
are adjusted to reproduce the free decay branches of 
the $\Lambda(1520)$.
Using Eq.~(\ref{VD})
the partial decay 
width of the $\Lambda(1520)$ into the $\bar{K}N$ or $\pi \Sigma$ 
decay channels can be calculated using the formula
\begin{eqnarray}
\label{LWidth}
\Gamma_{\scs 
\Lambda^*\to\Sigma(N)+\pi(\bar{K})}(s) &=& 
- 2 \mbox{Im}\Big[\Sigma(s)\Big]^{\Lambda^*}_{\Sigma \pi(\bar{K}N)}
\\
&=& \frac{1}{3} 
\left(\frac{g^2_{\scriptscriptstyle \Lambda^* \pi \Sigma(\Lambda^* \bar{K} N)} }{2 \pi}  \right) 
\frac{M_{\Sigma(N)}}{\sqrt{s}} |\kvec_{CM}|^5, \nonumber 
\end{eqnarray}
where
$|\kvec_{CM}| = 
{\lambda^{1/2}(s,M_{\Sigma(N)}^2,m_{\pi(\bar{K})}^2)}
/(2 \sqrt{s})
$ with
$s=P^2$ and $\lambda$ is the 
K\'alen function.
At the nominal pole position $P^2=M_{\Lambda(1520)} $ we get
$ {g}_{\scriptscriptstyle{\Lambda^* \pi \Sigma}} = 10.75~\mbox{GeV}^{-2}$ and
${g}_{\scriptscriptstyle{\Lambda^* \bar{K} N}} = 16.01~\mbox{GeV}^{-2}$. 
In non-dimensional limits, comparable to  
$g_{\scriptscriptstyle \Lambda^*\Sigma^*\pi}$ in Eq.~(\ref{TL})
\begin{equation}
[g^2_{\scriptscriptstyle \Lambda^* \pi \Sigma(\Lambda^* \bar{K} N)}
\, \kvec_{CM}^2]/3 \equiv 
\tilde{g}_{\scriptscriptstyle \Lambda^* \pi Y(\Lambda^* \bar{K} N)}
\end{equation}
the value of these couplings at the pole position of 
the $\Lambda^*(1520)$ would be
$\tilde{g}_{\scriptscriptstyle{\Lambda^* \pi \Sigma}} = 0.44$ and
$\tilde{g}_{\scriptscriptstyle{\Lambda^* \bar{K} N}}  = 0.54$,
which show that the  coupling
$g_{\scriptscriptstyle \Lambda^*\Sigma^*\pi}=0.89$
is still the largest one.

{\it $\Lambda^*(1520) \to \pi {\Sigma}(1189) $ channel}:
In the nuclear medium there are peculiarities which enforce us to 
consider the $\pi \Sigma$ and $\bar{K}N$ channels separately.
The $\pi \Sigma$-channel is a simplest one in the conventional 
decay of the $\Lambda^*(1520)$ hyperon.
The self energy loop integral in this decay channel reads
\begin{equation}
\label{SiSEF}
\Big[{\Sigma}(P,\rho)\Big]^{\Lambda^*}_{\pi {\it \Sigma}} = 
g^2_{\scriptscriptstyle{\Lambda^* \pi {\it\Sigma}}} \frac{i}{3}
\int \frac{d^4 k}{(2\pi)^4} \kvec^4 \tilde{D}_{\it \Sigma} (P-k,\rho) 
\tilde{D}_{\pi}(k,\rho)
\end{equation}
The imaginary part of Eq.~(\ref{SiSEF}) 
is meaningful by themselves
and can be obtained 
using the Cutkosky rules 
\begin{eqnarray}
\label{CR}
\Sigma(P,\rho) &\to& 2i \mbox{Im} \Big[ \Sigma(P,\rho) \Big], \nonumber \\
\tilde{D}_Y(K,\rho) &\to& 2i  
\mbox{Im} \Big[\tilde{D}_Y(K,\rho)\Big]  \cdot \theta(K^0), \nonumber\\
\tilde{D}_{\pi}(k,\rho) &\to&   2i \mbox{Im} \Big[\tilde{D}_{\pi}(k,\rho)\Big]  \cdot \theta(k^0). 
\end{eqnarray}
From this it is given by
\begin{widetext}
\begin{eqnarray}
\mbox{Im} 
\Big[
{\Sigma}(P,\rho)\Big]_{ \pi \it \Sigma}^{ \Lambda^*} &=&
 \frac{g^2_{\scriptscriptstyle{\Lambda^* \pi {\it\Sigma}}}}{3} 
\int \frac{d^3 \kvec}{(2\pi)^3}
\frac{\kvec^4 \, M_{\it \Sigma} }{E_{\it \Sigma}(\Pcapvec-\kvec)} \nonumber 
\\ && 
\left. 
\times \mbox{Im}
\Big[\tilde{D}_{\pi}(\omega,\kvec,\rho)\Big] \cdot \theta(\omega) \, 
\right|_{\ds \omega=P_0-E_{\it
  \Sigma}(\Pcapvec-\kvec)-V_{\it \Sigma}(\rho)}
\end{eqnarray}
\end{widetext}
where $V_{\it \Sigma}(\rho)$ is the binding correction 
for the $\Sigma(1189)$ hyperon, see Eq.~(\ref{BCorr}).

{\it $\Lambda^*(1520) \to \bar{K} N$ channel}: 
The proper treatment of the $\bar{K}N$ channel is a more subtle problem.
First, we consider the modification of the antikaon propagator 
in the nuclear medium. In its particle-antiparticle decomposition, the 
dispersion relation
representation of the $\bar{K}$ propagator is given by
\begin{equation}
\label{dressed_prop}
\tilde{D}_{\bar{K}} (k^0,\kvec;\rho) =
\int \limits_0^{\infty} d\, \omega 
\frac{S_{\bar{K}}
(\omega,\kvec;\rho)}{k^0-\omega+i0^+} - 
\int \limits_0^{\infty} d\, \omega
\frac{S_{K}
(\omega,\kvec;\rho)}{k^0+\omega-i0^+} 
\end{equation}
where  $S_{\bar{K} (K)}=-\mbox{Im}\tilde{D}_{\bar{K}(K)}/\pi$ is the spectral function of the $\bar{K} (K)$ meson
which 
depends on $\bar{K} (K)$ in-medium self energy $\Pi_{\bar{K} (K)}$ 
as provided
by Eqs.~(\ref{SF}) and~(\ref{PiSE}). As is well known, the 
interactions of $\bar{K}$ and $K$
with the nucleons of
the nuclear medium are rather different and, in principle, 
it is necessary to treat  them separately.

\begin{figure}[t]
\begin{center}
\includegraphics[clip=true,width=0.45\columnwidth,angle=0.]
{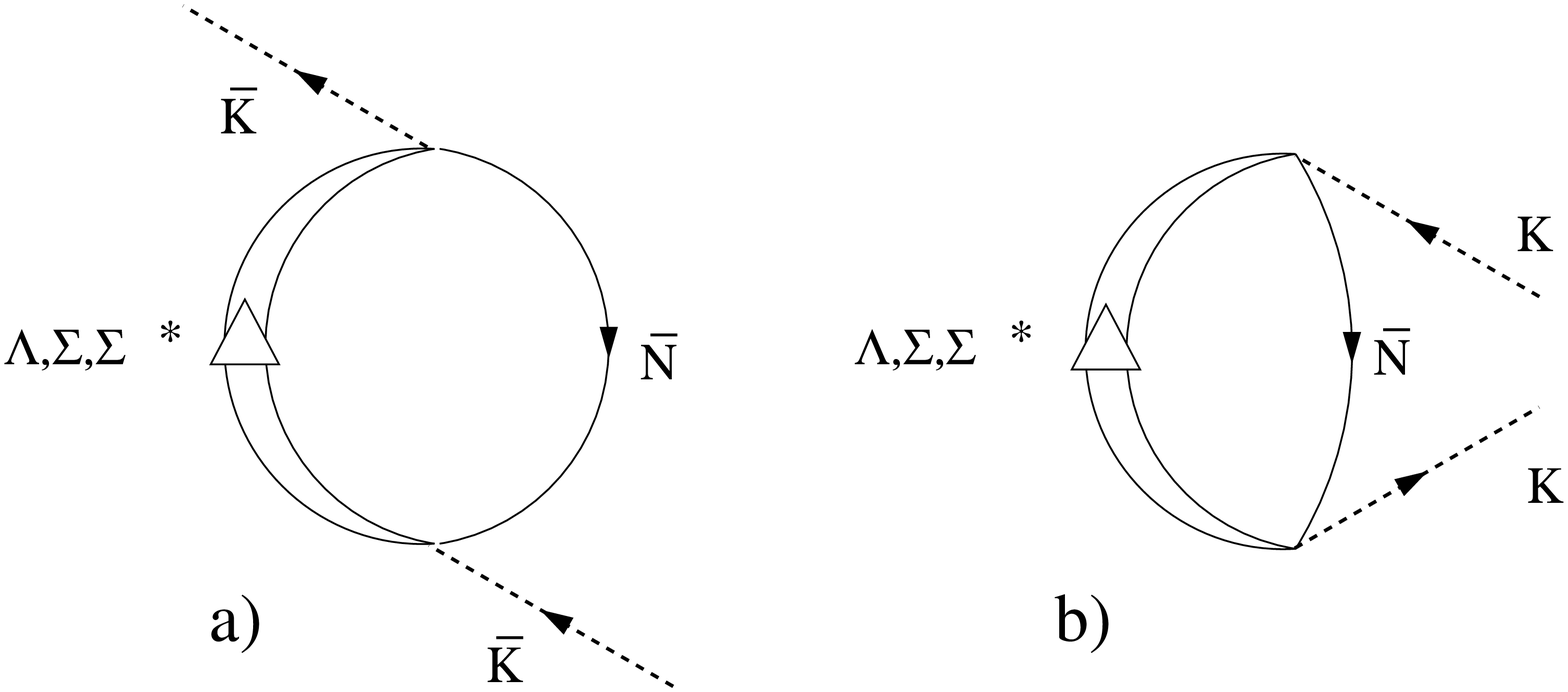}
\caption{\label{kaon_pwave} \footnotesize Kaon $P-$wave selfenergy diagrams: 
(a) $\bar{K}$ direct
term; (b) $K$ crossed term.}
\end{center}
\end{figure}

In the calculation of the in-medium selfenergy we need 
the in-medium nucleon propagator. It is given by
\begin{eqnarray}
\label{DN}
\tilde{D}_{N} (p,\rho) = \frac{M_N}{E_N(\pvec)} 
\frac{1-n(|\pvec|)}{P^0-E_N(\pvec)-V_N(\rho)+i0^+}
\nonumber \\
+ 
\frac{M_N}{E_N(\pvec)}  
\frac{n(|\pvec|)}{P^0-k^0-E_N(\pvec)-V_N(\rho)-i0^+}
\end{eqnarray}
where $n(|\pvec|)=\theta(k_F-|\pvec|)$ and $k_F$ is the 
Fermi momentum.
The first term in Eq.~(\ref{DN}) describes the Pauli blocked 
propagation of nucleons and the second
term is the hole propagator. The effect of the nucleon binding
is accounted for by the mean field potential
$V_N (\rho)\simeq -60 \rho/\rho_0$~MeV.
The selfenergy integral,
after the integration over the $k^0$ component takes the following form
\begin{widetext}
\begin{eqnarray}
\label{LKN}
\Big[\Sigma(P,\rho) \Big]^{ \Lambda^*}_{\bar{K} N} &=&  
{g}_{{\scriptscriptstyle \Lambda^* \bar{K} N}}^2 \frac{i}{3} 
\int \frac{d^4 k}{(2\pi)^4} \kvec^4 
\tilde{D}_{N} (P-k,\rho) 
\tilde{D}_{\bar{K}}(k,\rho) 
\nonumber \\
&=& - {g}_{{\scriptscriptstyle \Lambda^* \bar{K} N}}^2\frac{1}{3}  
\int \limits_0^{\infty} d\, \omega \int \frac{d^3 \kvec}{(2\pi)^3} 
\frac{M_N}{E_N(\Pcapvec-\kvec)} 
\frac{\kvec^4 [1-n(|\Pcapvec-\kvec|)]S_{\bar{K}}
(\omega,\kvec;\rho)}{P^0-E_N(\Pcapvec-\kvec)-\omega-V_N(\rho)+i0^+}
\nonumber \\
&& - {g}_{{\scriptscriptstyle \Lambda^* \bar{K} N}}^2 \frac{1}{3}   
\int \limits_0^{\infty} d\, \omega \int \frac{d^3 \kvec}{(2\pi)^3} 
\frac{M_N}{E_N(\Pcapvec-\kvec)} 
\frac{\kvec^4 n(|\Pcapvec-\kvec|) S_{{K}}
(\omega,\kvec;\rho)}{P^0-E_N(\Pcapvec-\kvec)+\omega-V_N(\rho)-i0^+}
\nonumber \\
\end{eqnarray}
In Eqs.~(\ref{LKN}) and~(\ref{SiSEF})  a static regulating form factor 
will be introduced in the results section.

Although the evaluation of Eq.~(\ref{LKN})
is straightforward,
in the present work we neglect the contribution
of the second term which gives no contribution to the imaginary part and only a
small one to the real part.
So we consider the spectral function of antikaons only. 
The corresponding imaginary part of the $\Lambda^*(1520)$ selfenergy 
is given by
\begin{eqnarray}
\mbox{Im} 
\Big[
{\Sigma}(P,\rho)\Big]_{\bar{K} N}^{\Lambda^*} &=&
-  g^2_{\scriptscriptstyle{\Lambda^* \bar{K} N}}  
 \frac{\pi}{3} 
\int \frac{d^3 \kvec}{(2\pi)^3}
\frac{\kvec^4 \, M_N }{E_N(\Pcapvec-\kvec)}  \\ 
&&\times [1-n(|\Pcapvec-\kvec|)] 
S_{\bar{K}} (\omega,\kvec,\rho)\cdot \theta(\omega) \, 
\Big|_{\ds \omega=P_0-E_N(\Pcapvec-\kvec)-V_N(\rho)} \nonumber
\end{eqnarray}
\end{widetext}
In the following  we shall briefly discuss how the antikaon spectral function 
is obtained. Here we follow closely Ref.~\cite{angelsk}.

The $P$-wave contribution to the ${\bar K}$ 
selfenergy comes from the coupling of the ${\bar K}$ meson to
hyperon particle-nucleon hole ($YN^{-1}$) excitations.
The corresponding many-body mechanisms are shown in Fig.~\ref{kaon_pwave}.
Because of strangeness conservation, only direct terms, 
Fig.~\ref{kaon_pwave}(a), are permitted for
the $\bar{K}$ excitations. Conversely, the $K$ selfenergy arises from the
crossed terms, Fig.~\ref{kaon_pwave}(b).
The $K^-$ meson can couple to $p\Lambda$, $p\Sigma^0$ or
$n\Sigma^-$ and the $\bar{K}^0$ to $n\Lambda$, $n\Sigma^0$ or
$p\Sigma^-$. The vertices $\bar{K}NY$ are derived from
the $D$ and $F$ terms of the chiral Lagrangian given in Appendix I
(see Eq.~(\ref{LagrangN})), expanding the unitary 
$SU(3)$ matrix $U$ up to one meson
field. Using a non-relativistic reduction of the
$\gamma^\mu\gamma^5$ matrix, one finds
\begin{eqnarray}
\label{PWBK}
-i t_{\bar{K}NY} &=& 
C_{\scriptscriptstyle \bar{K}NY}
(\sigmavec \cdot \kvec) \\
&=&
\left[\alpha_{\scriptscriptstyle \bar{K}NY} \frac{D+F}{2f} +
\beta_{\scriptscriptstyle \bar{K}NY} \frac{D-F}{2f}\right] (\sigmavec \cdot
\kvec)  \nonumber
\end{eqnarray}
where $\kvec$ is the incoming  $\bar{K}$ three momentum, $f=1.15 f_{\pi}$ and 
$\alpha_{\bar{K}NY}$,
$\beta_{\bar{K}NY}$ are the $SU(3)$ coefficients given in~\cite{angelsk}.

Following Ref.~\cite{Oset:2000eg} we consider 
the  $\bar{K}NY$ interaction in combination with the 
$\bar{K} N {{\Sigma}^*}(1385)$ transition where 
the expression for vertex function is given by
\begin{equation}
-it_{\bar{K}N\it{\Sigma}^*} = 
C_{\scriptscriptstyle \bar{K}N\it{\Sigma}^*} \,
(\Svec\,^\dagger\cdot\kvec) =
A_{\scriptscriptstyle \bar{K}N\it{\Sigma}^*} \,
\frac{2\sqrt{6}}{5} \frac{D + F}{2f}
(\Svec\, ^\dagger \cdot\kvec).
\label{eq:sigma}
\end{equation}
The $SU(3)$ coefficients $A_{\scriptscriptstyle \bar{K}N\it{\Sigma}^*}$ 
are given in  Ref.~\cite{Oset:2000eg}.
These couplings
were evaluated by first using the SU(6) quark model to relate the $\pi
N N$ coupling
to the $\pi N \Delta$ one and then
using SU(3) symmetry to
relate the $\pi N \Delta$
coupling to the $\bar{K} N \Sigma^*$ one, 
since the $\Sigma^*(1385)$
belongs to the SU(3) decuplet of the $\Delta$-isobar.

The $P$-wave $\bar{K}$ self energy in symmetric nuclear matter 
can then be summarized as
\begin{eqnarray}
\label{KPLF}
{\kvec}^2\,  \tilde{\Pi}^{(p)}_{\bar{K}}(\omega,\kvec,\rho) &=&
\frac{1}{2}  \left[ 
C^2_{\scriptscriptstyle K^- p \Lambda} 
\mathcal{R}_\Lambda^2 \right]  {\kvec}^2 \,
\mathcal{U}_\Lambda(\omega,\kvec,\rho) \nonumber \\
&+& \frac{3}{2} \left[ 
{C}^2_{\scriptscriptstyle K^- p {\it{\Sigma}}^0}
 \mathcal{R}_\Sigma^2\right] {\kvec}^2 \, 
\mathcal{U}_\Sigma(\omega,\kvec,\rho)  \\
&+& \frac{1}{2} \left[ 
{C}^2_{\scriptscriptstyle K^- p {\it{\Sigma}}^{*0}}
 \mathcal{R}_{\Sigma^*}^2 \right] 
{\kvec}^2 \, \mathcal{U}_{\Sigma^*}(\omega,\kvec,\rho) \nonumber 
\label{eq:selfkap}
\end{eqnarray}
where the Lindhard function $\mathcal{U}_Y(q)$ 
$(Y=\Lambda,\Sigma$ or $\Sigma^*$) 
accounts for the direct term only,
see Fig.~\ref{kaon_pwave} (a). Its explicit expression can be found in
Ref.~\cite{Oset:2000eg}.
In Eq.~(\ref{KPLF}) 
$\mathcal{R}_{\Lambda(\Sigma)}=(1-\omega/2M_{\Lambda(\Sigma)})$, 
$\mathcal{R}_{\Sigma^*}=(1-\omega/M_{\Sigma^*})$ are the
relativistic recoil vertex corrections~\cite{angelsk}.
In addition, we use the static form-factors at 
the antikaon-baryon vertices of monopole type, $\Lambda^2/(\Lambda^2+\kvec^2)$,
with $\Lambda=1$ GeV. 

Finally, we take into account the short-range correlations in the hyperon-hole 
$YN^{-1}-YN^{-1}$ channels using a standard prescription
\begin{equation}
\Pi_{\bar{K}}(\omega,\kvec,\rho) 
= \frac{\kvec^2 \tilde{\Pi}^{(p)}_{\bar{K}}(\omega,\kvec,\rho)}
{1-g'\, \tilde{\Pi}^{(p)}_{\bar{K}}(\omega,\kvec,\rho)}
\end{equation}
where we assume the same value of the Landau-Migdal parameter 
$g'=0.7$ as in Eq.~(\ref{PolOperator}).

\begin{figure*}[t]
\begin{center}
\includegraphics[clip=true,width=0.7287\columnwidth,angle=0.]
{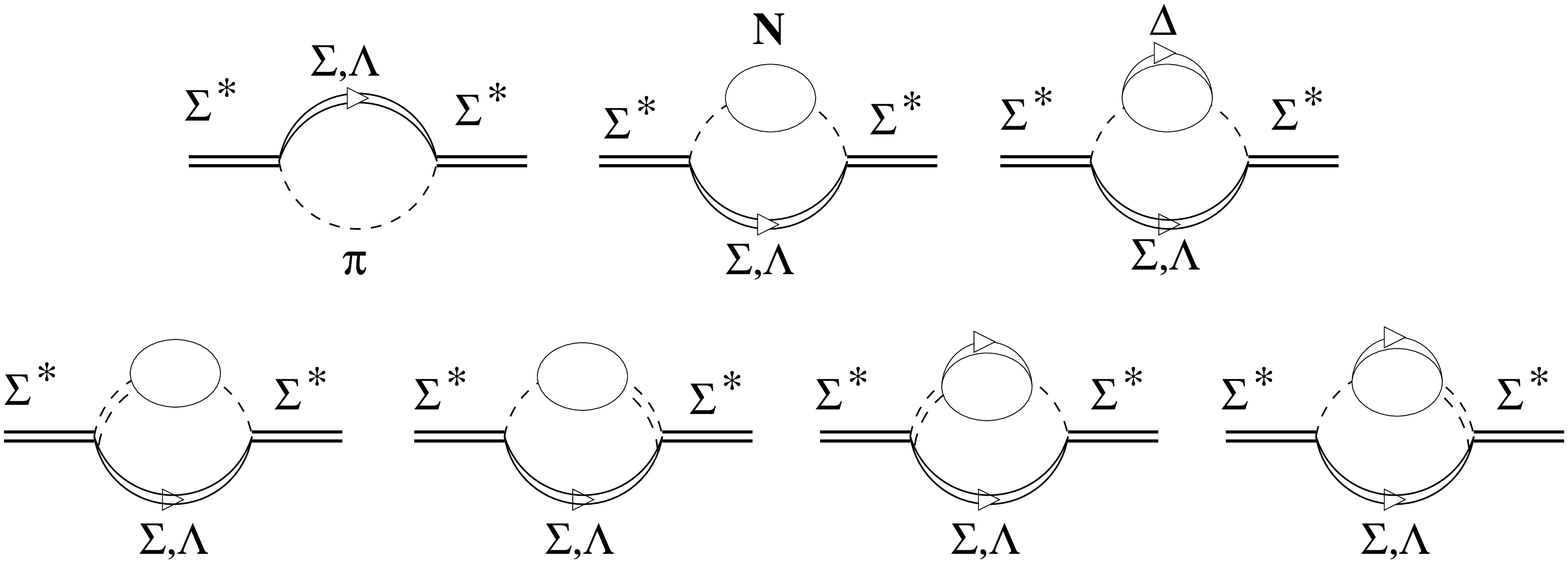}
\caption{\label{RenormSigma} \footnotesize 
In-medium renormalization of the $\Sigma^*(1385)$ in the 
$\pi \Sigma + \pi \Lambda$ channels. The last four
diagrams account for the short-range correlations.}
\end{center}
\end{figure*}

\section{\label{S2} Renormalization of the $\Sigma^*(1385)$}

As we have seen from Eq.~(\ref{SELambda}), 
the problem of the in-medium modification of the $\Lambda(1520)$ 
in the $\pi \Sigma^*(1385)$ channel can be 
reduced to the proper description of the properties of pions 
and $\Sigma^*(1385)$ in the nuclear medium.
The renormalization scheme which we
employ for the $\Sigma^*(1385)$ is essentially
the same as in the previous case except for the $P$-wave nature of 
the hadronic $\Sigma^*(1385)$ decay. This implies some peculiarities, for
instance, the proper treatment of short range correlations.

{\it $\Sigma^*(1385) \to \pi \Lambda + \pi \Sigma$ channel} :
We consider the following $P$-wave decay branches
of the $\Sigma^*(1385)$ hyperon:
$\Sigma^*(1385) \to \Lambda(1115) + \pi$ and 
$\Sigma^*(1385) \to \Sigma(1189) + \pi $.
In what follows we shall further
dress the octet states $\Lambda(1115)$ and  
$\Sigma(1189)$ using the phenomenological optical potentials.
The vertex functions describing the transitions are given by
\begin{equation}
-i t_{\pi Y \to \Sigma^*} = 
- C_{\scriptscriptstyle  \pi Y {\it\Sigma}^*} (\Svec^{\dagger} \cdot \kvec)
\end{equation}
where $Y=\Sigma(1189)$
or $\Lambda(1115)$. 
For coupling constants $C_{\scriptscriptstyle \pi Y{\it\Sigma}^*}$
we follow Ref.~\cite{Oset:2000eg} 
and use the quark model values
\begin{equation}
\label{C}
C_{\scriptscriptstyle \pi \Lambda {\it\Sigma}^*} 
= \frac{6}{5} \frac{D+F}{2 f_{\pi}},~~~~~
C_{\scriptscriptstyle \pi {\it \Sigma} {\it\Sigma}^*} 
= -\frac{2\sqrt{3}}{5} \frac{D+F}{2 f_{\pi}} 
\end{equation}
From this,  the explicit expressions for 
$\Sigma_{\pi Y}^{{\it\Sigma}^*}$ are given by
\begin{eqnarray}
\label{Si1}
\Big[\Sigma(P,\rho)\Big]_{\pi \Lambda(1115)}^{{\it \Sigma}^*(1385)} = 
i \left( C_{\scriptscriptstyle \pi \Lambda {\it\Sigma}^*}
\right)^2 
\frac{1}{3}
\int \frac{d^4 k}{(2 \pi)^4}
\kvec^2
\tilde{D}_{\Lambda}(P-k,\rho) \tilde{D}_{\pi}(k,\rho)
\end{eqnarray}
\begin{eqnarray}
\label{Si2}
\Big[ \Sigma(P,\rho) \Big]_{\pi {\it\Sigma}(1189)}^{{\it\Sigma}^*(1385)} = 
i \left( C_{\scriptscriptstyle \pi {\it \Sigma} {\it\Sigma}^*} \right)^2 
\frac{2}{3}
\int \frac{d^4 k}{(2 \pi)^4}
\kvec^2
\tilde{D}_{\it \Sigma}(P-k,\rho) \tilde{D}_{\pi}(k,\rho)
\end{eqnarray}
The additional factor 2 in Eq.~(\ref{Si2}) comes from contribution of two 
possible charge states.

At this point we would like to mention that in a realistic calculation one
would have to add strong repulsive forces at short distances 
which would generate
short-range correlation.  
The correlations of this type of the interaction would 
effectively modulate the in-medium $\pi$ exchange 
interaction~\cite{Oset:1979bi,Kaskulov:2005kr}, introducing the 
correlation parameter $g'$. 
The denominator in 
Eq.~(\ref{PolOperator})
takes into account this effect between $P$-wave bubbles in the diagrams 
of Fig.~\ref{RenormSigma}, but
not between the external hyperon and the contiguous bubble. To account for
this
we make the separation between the longitudinal $\mathcal{V}_l$ and 
transverse $\mathcal{V}_t$ parts of the
pion effective interaction~\cite{Alberico:1981sz,Oset:1987re}
in the $P$-wave loop integrals
\begin{eqnarray}
\left(\frac{D+F}{2 f_{\pi}}\right)^2 
\frac{k_i k_j}{(k^{0})^2-\kvec^2-m_{\pi}^2 + i0^+} 
 \longrightarrow \mathcal{V}_l(k) \, \hat{k}_i  \hat{k}_j 
+ \mathcal{V}_t(k) (\delta_{ij} - \hat{k}_i  \hat{k}_j) 
\end{eqnarray}
Here $k_i$ is the Cartesian component of the unit vector 
$\hat{\kvec}=\kvec/|\kvec|$ and
\begin{eqnarray}
\label{Vl}
\mathcal{V}_l(k) &=& \left(\frac{D+F}{2 f_{\pi}}\right)^2 
\left[ \frac{\kvec^2}{(k^{0})^2-\kvec^2-m_{\pi}^2 + i 0^+}  + g'\right]
F(\kvec)^2 \nonumber \\
\label{Vt}
\mathcal{V}_t(q) &=& \left(\frac{D+F}{2 f_{\pi}}\right)^2 \, g' \, F(\kvec)^2 
\end{eqnarray}
This procedure can be represented by the last four diagrams in 
Fig.~\ref{RenormSigma}.
In Eqs.~(\ref{Vl}) and~(\ref{Vt}) $F(\kvec)$ is a static form factor
$\Lambda^2/(\Lambda^2+\qvec^2)$ with the cut off scale $\Lambda=1$~GeV.
 Hence, we 
make the following substitution in the selfenergy 
of the $\Sigma^*(1385)$, see Eqs.~(\ref{Si1})
and~(\ref{Si2})
\begin{eqnarray}
\left(\frac{D+F}{2 f_{\pi}}\right)^2 \kvec^2 \tilde{D}_{\pi}(k,\rho) \to 
\mathcal{W}(k,\rho) 
= \frac{\mathcal{V}_l(k)}{1- \mathcal{U}(k,\rho) \mathcal{V}_l(k)}
+ \frac{2 \, \mathcal{V}_t(k)}{1- \mathcal{U}(k,\rho) \mathcal{V}_t(k)}.
\end{eqnarray}
Using the Cutkosky rules 
Eqs.~(\ref{CR}) supplemented by
\begin{eqnarray}
\mathcal{W}(k,\rho) &\to& 2i   
\mbox{Im}\Big[\mathcal{W}(k,\rho)\Big] \cdot \theta(k^0)
\end{eqnarray}
 one may calculate
the imaginary part of the loop  integrals which are given by
\begin{widetext}
\begin{eqnarray}
\mbox{Im}\Big[\Sigma(P,\rho)\Big]_{\pi \Lambda}^{\Sigma^*} = 
\left(\tilde{C}_{\scriptscriptstyle \pi \Lambda{\it\Sigma}^*}\right)^2 
\frac{1}{3} 
\int \frac{d^3 \kvec}{(2\pi)^3}
\frac{M_{\Lambda}}{E_{\Lambda}(\Pcapvec-\kvec)} \nonumber 
\left. 
\mbox{Im}
\Big[\mathcal{W}(k,\rho) \Big]\cdot \theta(k^0) \, \right|_{\ds
k^0=P^0-E_{\Lambda}(\Pcapvec-\kvec)} 
\end{eqnarray}
\begin{eqnarray}
\mbox{Im}\Big[\Sigma(P,\rho)\Big]_{\pi \Sigma}^{\Sigma^*} = 
\left(\tilde{C}_{\scriptscriptstyle \pi 
{\it\Sigma}{\it\Sigma}^*}\right)^2 \frac{2}{3} 
\int \frac{d^3 \kvec}{(2\pi)^3}
\frac{M_{\Sigma}}{E_{\Sigma}(\Pcapvec-\kvec)} \nonumber 
\left.   \mbox{Im}
\Big[\mathcal{W}(k,\rho) \Big]\cdot \theta(k^0) \, \right|_{\ds
k^0=P^0-E_{\Sigma}(\Pcapvec-\kvec)} 
\end{eqnarray}
\end{widetext}
where $\theta$ is the step function and 
$\tilde{C}_{\scriptscriptstyle \pi Y{\it\Sigma}^*}$  are the 
reduced coupling constants 
obtained from Eq.~(\ref{C}) by omitting the
factor $(D+F)/2f_{\pi}$.

The vacuum subtracted expression for 
$\mathcal{W}(k,\rho)$ reads
\begin{eqnarray}
\delta \mathcal{W}(k,\rho) &=& \mathcal{W}(k,\rho) - \mathcal{W}(k,0) 
= 
\frac{\mathcal{U}(k,\rho) \mathcal{V}_l^2(k)}{1- \mathcal{U}(k,\rho) \mathcal{V}_l(k)}
+ \frac{2\, \mathcal{U}(k,\rho) \mathcal{V}_t^2(k)}{1- \mathcal{U}(k,\rho) 
\mathcal{V}_t(k)}
\end{eqnarray}
From this the  subtracted versions of the selfenergy integrals
Eqs.~(\ref{Si1}) and~(\ref{Si2})  take the form
\begin{eqnarray}
\label{Si1W}
\Big[ \delta \Sigma (P,\rho)\Big]^{\Sigma^*}_{\pi \Lambda} = 
i \left(\tilde{C}_{\scriptscriptstyle \pi {\it \Sigma} {\it\Sigma}^*} \right)^2 \frac{1}{3}
\int \frac{d^4 k}{(2 \pi)^4}
\tilde{D}_{\Lambda}(P-k,\rho) \, \delta \mathcal{W}(k,\rho) \\
\label{Si2W}
\Big[\delta \Sigma (P,\rho)\Big]^{\Sigma^*}_{\pi \Sigma} = 
i \left(\tilde{C}_{\scriptscriptstyle \pi \Lambda{\it\Sigma}^*} \right)^2 
\frac{2}{3} \int \frac{d^4 k}{(2 \pi)^4}
\tilde{D}_{\Sigma}(P-k,\rho) \, \delta \mathcal{W}(k,\rho)
\end{eqnarray}
Using Eqs.~(\ref{Si1W}) and~(\ref{Si2W}) we can evaluate the in-medium
modification of both the real
and imaginary parts. We also take into account the phenomenological 
binding corrections
to $\Sigma$ and $\Lambda$ in nuclei to which the $\Sigma^*(1385)$ decays.
These corrections are accounted for by~\cite{Oset:1989ey}
\begin{equation}
\label{BCorr}
V_{\Lambda}(\rho)=\mbox{Re} \Sigma_{\Lambda}= 
V_{\it \Sigma}(\rho)=\mbox{Re} \Sigma_{\it \Sigma}=-30 \rho/\rho_0
\end{equation}

\begin{figure*}[t]
\begin{center}
\includegraphics[clip=true,width=0.99\columnwidth,angle=0.]
{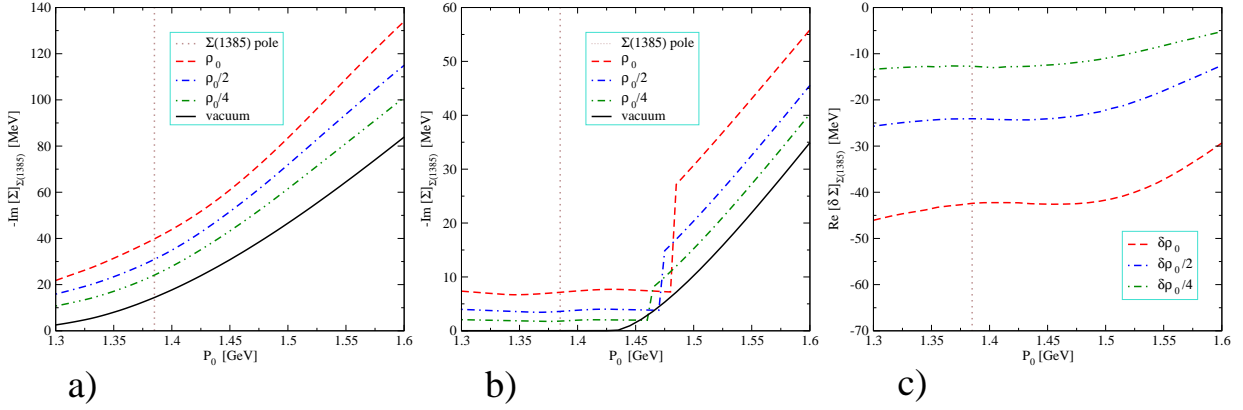}
\caption{\label{Sigma1385} \footnotesize  
The imaginary parts (a) and (b) of the $\Sigma^*(1385)$ selfenergy 
in the $(\pi \Sigma + \pi \Lambda)$ and $\bar{K} N$ channels, respectively,
at several
  densities $\rho=\rho_0$ (dashed curve), $\rho_0/2$ (dot-dashed curve) and 
$\rho_0/4$ (dot-dot-dashed curve) as a function of 
$P^{\mu}_{\Sigma^*(1385)}=(P^0,0)$.
The vacuum
  subtracted real  part (c) 
of the $\Sigma^*(1385)$ selfenergy.
The dotted vertical line indicates the  $\Sigma^*(1385)$ pole position.
}
\end{center}
\end{figure*}

{\it $\Sigma^*(1385) \to \bar{K} N$ channel} :
The decay modes considered in the previous sections are allowed in the free 
space. But there are channels like $\Sigma^*(1385) \to \bar{K} N$ 
which may open up in the medium because of the additional phase space. 
Using the vertex function given by Eq.~(\ref{eq:sigma})
the expression for the selfenergy integral reads
\begin{widetext}
\begin{eqnarray}
\label{Si3}
\Big[\Sigma(P,\rho)\Big]_{\bar{K} N}^{{\it \Sigma}^*(1385)} = 
\left( C_{\scriptscriptstyle \bar{K} N {\it\Sigma}^*}
\right)^2 
\frac{i}{3} 
\int \limits_0^{\infty} d\, \omega 
\int \frac{d^4 k}{(2 \pi)^4}
\kvec^2
\tilde{D}_{N}(P-k,\rho) 
\left[\frac{S_{\bar{K}}
(\omega,\kvec;\rho)}{k^0-\omega+i0^+} - 
\frac{S_{K}
(\omega,\kvec;\rho)}{k^0+\omega-i0^+} \right] \nonumber \\
\end{eqnarray}
\end{widetext}
where $\tilde{D}_N$ is the in-medium nucleon 
propagator, Eq.~(\ref{DN}), and  $\tilde{D}_{\bar{K}}$ is the $\bar{K}$ 
propagator which is introduced in accord with Eq.~(\ref{dressed_prop}). 
The direct application of that result is not entirely
correct because we deal with the $P$-wave kaons which are also affected by the
short range correlations in this channel. For instance, in the expression for
the imaginary part of the in-medium selfenergy we obtain
\begin{widetext}
\begin{eqnarray}
\mbox{Im} 
\Big[
{\Sigma}(P,\rho)\Big]_{\bar{K} N}^{\Sigma^*} &=&
(\tilde{C}_{ \scriptscriptstyle{\bar{K} N \Sigma^* }})^2
 \frac{1}{3} 
\int \frac{d^3 \kvec}{(2\pi)^3}
\frac{M_N }{E_N(\Pcapvec-\kvec)}  \\ 
&&\times [1-n(|\Pcapvec-\kvec|)] 
\mbox{Im}\Big[\mathcal{W}_{\bar{K}}(\omega,\kvec,\rho)\Big] (\omega,\kvec,\rho)\cdot \theta(\omega) \, 
\Big|_{\ds \omega=P_0-E_N(\Pcapvec-\kvec)-V_N(\rho)} \nonumber
\end{eqnarray}
\end{widetext}
where $\tilde{C}_{ \scriptscriptstyle{\bar{K} N \Sigma^* }} =
{C}_{ \scriptscriptstyle{\bar{K} N \Sigma^* }} (2f_{\pi})/(D+F)$
and
\begin{equation}
\label{WKbar}
\mathcal{W}_{\bar{K}} = 
\frac{\mathcal{V}_{l,\bar{K}}(k)}{1- \mathcal{U}_{\bar{K}}(k,\rho) 
\mathcal{V}_{l,\bar{K}}(k)}
+ \frac{2 \, \mathcal{V}_{t,\bar{K}}(k)}{1- \mathcal{U}_{\bar{K}}(k,\rho) 
\mathcal{V}_{t,\bar{K}}(k)}
\end{equation}
with $\mathcal{V}_l$ and $\mathcal{V}_t$ given by Eq.~(\ref{Vt}) replacing
$m_{\pi}$ by $m_{\bar{K}}$ and using the same $g'$. In addition,
in Eq.~(\ref{WKbar})  $ \mathcal{U}_{\bar{K}} = \tilde{\Pi}^{(p)}_{\bar{K}}
/(\frac{D+F}{2 f_{\pi}})^{2}$.

\section{Results for the $\Sigma^*(1385)$}
Our results for the imaginary part of the
$\Sigma^*(1385)$, 
or equivalently the 
width $\Gamma_{\Sigma^*} = - 2 \mbox{Im} \Sigma_{\Sigma^*}$
in the $\pi \Sigma + \pi \Lambda$ decay channels
and in the reference
frame where $P=(P_0,0)$ are shown in Fig.~\ref{Sigma1385}~(a) for several 
densities. 
The vacuum value at the nominal pole position 
is $\Gamma_{\Sigma^*}\simeq 30$~MeV
and in good agreement with its empirical value 
$\Gamma_{\Sigma^*}=35 \pm 4$~\cite{Groom:2000in}. Also the vacuum branching
ratio $\Gamma_{\Lambda}/\Gamma_{\Sigma}\simeq 7.7$ compares well with
experiment $7.5 \pm 0.5$. Increasing the density
we observe the broadening of the $\Sigma^*(1385)$ hyperon, and as a result, 
at normal nuclear matter densities $\rho_0=0.16$~fm$^{-3}$
the width becomes $\simeq 76$~MeV. Note, that the SR
correlations play here an important role. We find that, 
without SR correlations 
the effect of the medium on $\Sigma^*(1385)$  is 
unrealistically big and produces the increase of the width relative to 
the vacuum
value by a factor five. This situation is similar to the  
behavior of the $\Delta(1232)$-isobar at finite density where the SR 
correlations play an important role 
and strongly moderate the change of mass and width of the
$\Delta$ isobar~\cite{Oset:1987re}. 

The results for another related channel, namely $\Sigma^*(1385) \to \bar{K} N$
is shown in Fig.~\ref{Sigma1385}~(b). Because of the Pauli blocking
and relatively weak interaction of $\bar{K}$  with the nuclear medium 
as compared with pions 
the impact of this channel is small and adds an additional
portion $\simeq 7$~MeV to the imaginary part. This is much smaller than
obtained in Ref.~\cite{Lutz:2001dq} where only the $\bar{K}N$ channel is considered
and an on-shell approximation of the $P$-wave $\bar{K}N$ amplitude is done.
This approximation is not good for nuclei since it ties the momentum
to the energy of the $\bar{K}$, which becomes imaginary below the 
$\bar{K}N$ threshold. However the $k^0$ and $\kvec$ variables in matter are
independent variables and $\kvec$ is always physical.

For the 
real part of the $\Sigma^*(1385)$ self energy - the in-medium mass shift -
we find an attractive potential at normal nuclear matter
density with a strength of about $\simeq - 45$~MeV,
see Fig.~\ref{Sigma1385}~(c).

\begin{figure*}[t]
\begin{center}
\includegraphics[clip=true,width=0.99\columnwidth,angle=0.]
{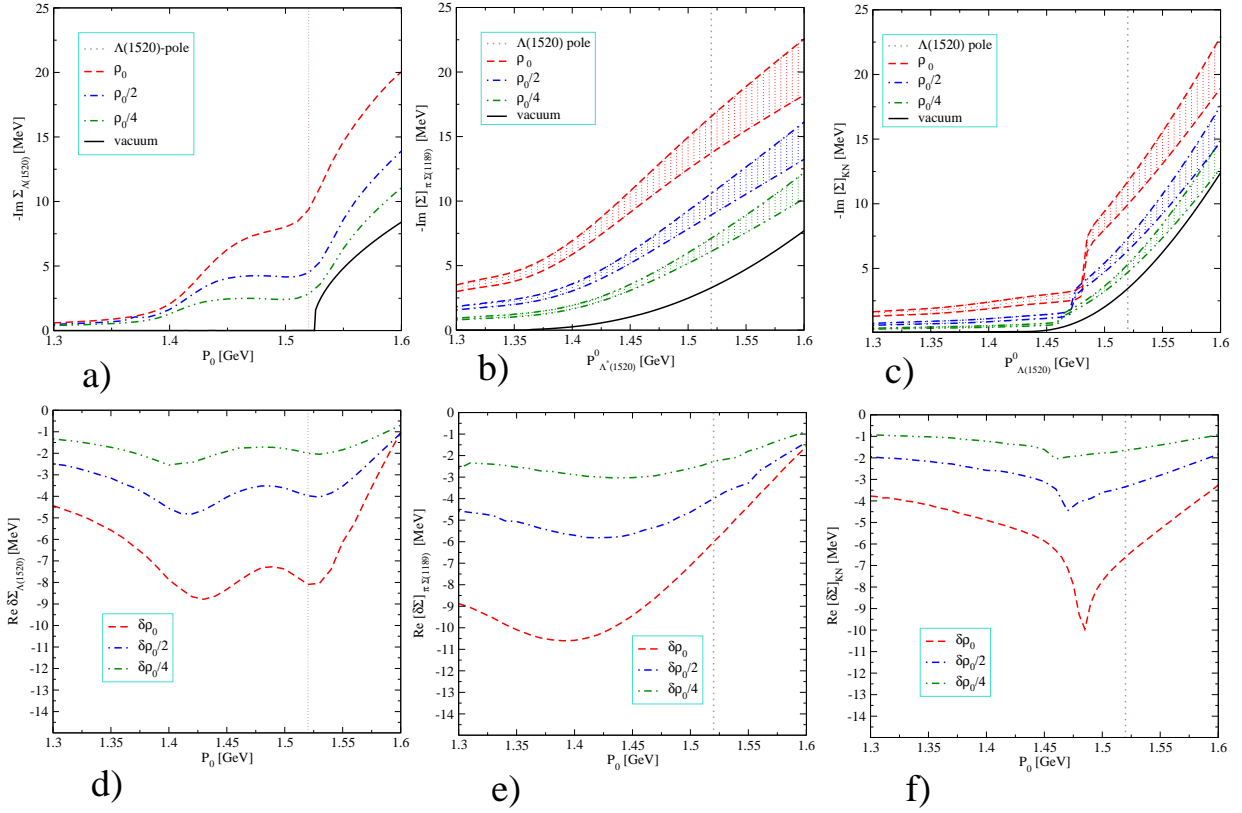}
\caption{\label{RenormLKbarN} \footnotesize 
In-medium renormalization of the $\Lambda(1520)$ in the $\pi \Sigma^*(1385)$~a,d)
$\pi \Sigma$~b,e) and $\bar{K}N$~c,f)
channels. The top and bottom panels are the imaginary and 
the vacuum subtracted real parts of the selfenergies, respectively.
The vertical line indicates the  $\Lambda(1520)$ pole position.
}
\end{center}
\end{figure*}

\begin{figure*}[t]
\begin{center}
\includegraphics[clip=true,width=0.99\columnwidth,angle=0.]
{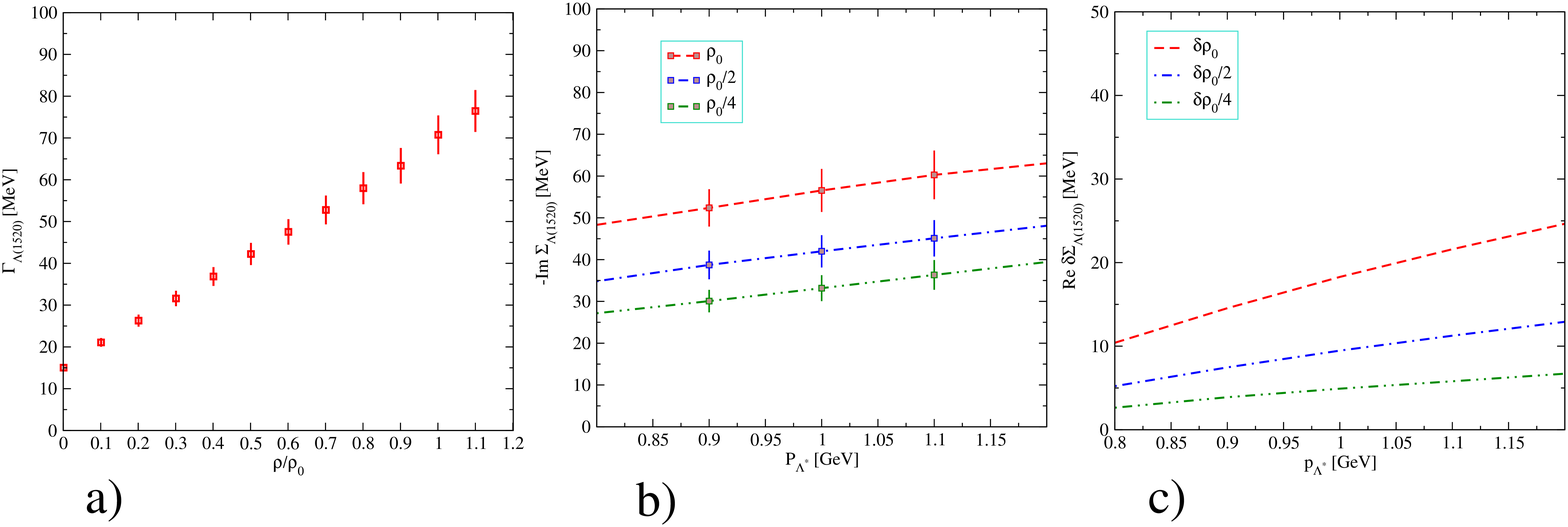}
\caption{\label{LambdaPlab} \footnotesize 
Values with theoretical uncertainties for the width of the $\Lambda(1520)$
at rest in the medium, including the free width, 
as function of the nuclear matter density
$\rho/\rho_0$ (a).
The 
imaginary (b) and real (c) parts of the
$\Lambda(1520)$ selfenergy as a function of a three 
momentum $|\Pcapvec_{\Lambda}|$.
}
\end{center}
\end{figure*}

\section{Results for the $\Lambda^*(1520)$}

Using the dressed $\Sigma^*(1385)$ discussed in the previous
section we
calculate  the selfenergy of the $\Lambda^*(1520)$ given by 
Eq.~(\ref{SELambda}) in the $\pi \Sigma^*(1385)$ decay channel. 
Our results for the in-medium mass and width of the $\Lambda(1520)$ 
in this novel $\pi \Sigma^*$ channel are 
shown in
Fig.~\ref{RenormLKbarN}~(a,d). 
As we have already noted 
for the nominal masses of hyperons involved there is an 
energy gap $\simeq 5$~MeV which makes 
the decay $\Lambda(1520) \to \pi \Sigma^*(1385)$ not possible. 
In the nuclear medium the pions decay to $p-h$ and $\Delta-h$
and this effect allows to open the
$\Lambda(1520) \to (ph) \Sigma^*(1385)$ decay channel which has a large
available phase space. 
In Fig.~\ref{RenormLKbarN} the dashed, dot-dashed and dot-dot-dashed
curves correspond to $\rho_0$,
$\rho_0/2$ and $\rho_0/4$, respectively. In this channel 
we get for the width of the $\Lambda(1520)$ 
$\Gamma_{\Lambda(1520)}\simeq 18$~MeV at the nominal pole position
and normal nuclear matter density.
This value is even bigger than the free width of the 
$\Lambda(1520)$ which is $\Gamma_{\Lambda(1520)}\simeq 15$~MeV.
The corresponding results for the vacuum subtracted real part 
of the selfenergy  are shown 
if Fig.~\ref{RenormLKbarN}~(d). Here we 
find relatively weak attractive potential of about
$\simeq -8$~MeV at the resonance pole position and 
normal nuclear matter density.

The renormalization of the $\Lambda^*(1520)$ in 
the $\pi \Sigma(1189)$ channel is shown in Fig.~\ref{RenormLKbarN}~(b,e). 
These curves correspond to the regularization of 
the selfenergy integral Eq.~(\ref{SiSEF}) when using the form factor
$[F(\kvec^2)]^2=[\Lambda^2/(\Lambda^2+\kvec^2)]^2$ with
the cut off scale $\Lambda=450\div 500$~MeV which is needed for
$D$-wave loops in Ref.~\cite{RSO}.
We normalize the form factor to unity at the $\Lambda^*(1520)$ 
pole position $[F(\kvec^2)/F(\kvec_{on-shell}^2)]^2$ where $\kvec_{on-shell}$ 
is the on-mass-shell three momenta of the meson in the loop.
 We use both limiting values and consider them as a
sort of theoretical uncertainties. The results are presented as a band where
the upper limit correspond to the $\Lambda = 500$~MeV and the lower limit
to the $\Lambda = 450$~MeV. For instance, for the $\Lambda = 500$~MeV we 
get $\simeq 32$~MeV width of the $\Lambda^*(1520)$ at normal nuclear matter 
density $\rho_0$ and for $\Lambda=450$~MeV we get $\simeq 26$~MeV.  
One should compare this result with the free decay in 
this channel (solid curve) where the corresponding value is $\simeq 7$~MeV 
only. 
The vacuum subtracted real part of the selfenergy is shown in 
Fig.~\ref{RenormLKbarN}~(e) and was calculated with cut off 
$\Lambda = 500$~MeV. The changes
are moderate and for $\rho=\rho_0$ we get the attraction  $\simeq -6$~MeV 
at energies near the  pole position.
In Fig.~\ref{RenormLKbarN}~(c,f) we show our results for 
the renormalization of the $\Lambda(1520)$ in the 
$\bar{K} N$ channel. Here the 
results are
quantitatively  similar
to the $\pi \Sigma$ channel. At normal density we get the width
$\simeq 20$~MeV  and additional
attraction $\simeq -7$~MeV.
With present uncertainties 
we give a band of values for 
the width of the $\Lambda(1520)$ in the
nuclear medium including now the free width and the in-medium renormalization
from the $\pi \Sigma^*(1385)$, $\bar{K}N$ and $\pi \Sigma$ related channels. 
We show these results in Fig.~\ref{LambdaPlab}~(a) 
as a function of $\rho/\rho_0$.
As one can see, at $\rho=\rho_0$ we get a $\Lambda(1520)$ width of about
$\simeq 70\div 80$~MeV, which is about five times the free width.
These results are of the same order of magnitude as those obtained
in Ref.~\cite{Lutz:2001dq} with the $\bar{K}N$ channel alone. The comments
made above about the approximations done in~\cite{Lutz:2001dq} hold also
in the present case.

Finally, we extend the discussion to some particular kinematic
relevant for possible application of the presented formalism
to reactions like $\gamma p \to K^+ \Lambda(1520)$ in nuclei where the
$\Lambda(1520)$ hyperon is produced with large momentum. In 
Fig.~\ref{LambdaPlab} the imaginary (b) and real (c) 
parts of the selfenergy are shown for the
$\Lambda(1520)$ moving in nuclear matter as a function of a three 
momentum $|\Pcapvec_{\Lambda}|$. We can see that the imaginary part of the
$\Lambda(1520)$ selfenergy is not changed much from its value at zero
momentum. However, the real part changes sign from $|\Pcapvec_{\Lambda}|=0$ to 
$|\Pcapvec_{\Lambda}| \simeq 1000$~MeV. But in both cases these changes are 
relatively small. Even if the experiment quoted above would be most suited
to determine the $\Lambda(1520)$ width in the nucleus, there is already
experimental information which allows us to get some hint on its size.
The study performed in heavy ion collisions~\cite{R1,R2} indicates that
a better agreement of theory with experimental data is obtained assuming that
about half of the $\Lambda(1520)$ produced are absorbed in the nucleus.
Such a reduction can only be obtained with an in-medium width of tens of MeV 
as one can guess from comparison to studies done in $\phi$ production
in the $pA$ reaction~\cite{Magas:2004eb}.

\section{Conclusions}
We have addressed the problem of the self energy of the $\Lambda(1520)$ and 
$\Sigma^*(1385)$ resonances in a nuclear medium and we have found relevant 
changes in the medium in both the real and imaginary parts, particularly in
the latter. Considering the coupled channel character of the $\Lambda(1520)$
resonance, where the $\pi \Sigma^*$, $\bar{K}N$ and $\pi \Sigma$ channels play
a very important role, particularly the $\pi \Sigma^*$, one finds pronounced
changes in the width in the nuclear medium when the $\pi$ component is allowed
to become a $p-h$ excitation and the $\bar{K}$ 
component a $\Lambda-h$, $\Sigma-h$
or $\Sigma^*-h$ excitations. All the three channels when these $p-h$ or 
$Y-h$ excitations are allowed, increase the width by about $15\div 20$~MeV, a 
larger amount than the free width, as a consequence of which one obtains at
the end a $\Lambda(1520)$ width at $\rho =\rho_0$ of about four to five
times the free width.

Such a spectacular change should be in principle easily observable
experimentally. It will be interesting to look at suitable reactions to
measure this. Recently there have been interesting developments in this
direction and experiments to measure changes in the $\phi$ width in the medium
have been conducted~\cite{Ishikawa:2005aw} or are being proposed~\cite{Hartman}
by looking at the $A$ dependence of the $\phi$ production in nuclei.
Theoretical calculations of this $A$ dependence show indeed that the method
is suited and offers advantage over other 
methods~\cite{Cabrera:2003wb,Magas:2004eb}. 
The investigation of these
medium effects would be a novel and interesting enterprise which would shed
much information of the nature of the $\Lambda(1520)$ resonance.

\section*{Acknowledgments}
This work is partly supported by DGICYT contract number BFM2003-00856,
and the E.U. EURIDICE network contract no. HPRN-CT-2002-00311.
This research is part of the EU Integrated Infrastructure Initiative
Hadron Physics Project under contract number RII3-CT-2004-506078.

\begin{appendix}
\section{Effective $SU(3)$ Lagrangian}
The lowest order chiral
Lagrangian,
coupling the octet of pseudoscalar mesons to the octet of $1/2^+$
baryons, is
\begin{eqnarray}
\label{LagrangN}
L_1^{(B)} = &&\langle \bar{B} i \gamma^{\mu} \nabla_{\mu} B
\rangle -
M_B \langle \bar{B} B \rangle \\
& + &\frac{1}{2} D \langle \bar{B} \gamma^{\mu} \gamma_5 \left\{
u_{\mu},
B \right\} \rangle
+ \frac{1}{2} F \langle \bar{B} \gamma^{\mu} \gamma_5 [u_{\mu},
B]
\rangle \nonumber
\end{eqnarray}
where 
$\langle\, \rangle$ denotes the trace of SU(3)
matrices
and
\begin{equation}
\begin{array}{l}
\nabla_{\mu} B = \partial_{\mu} B + [\Gamma_{\mu}, B] \\
\Gamma_{\mu} = \frac{1}{2} (u^\dagger \partial_{\mu} u + u
\partial_{\mu} u^\dagger) \\
U = u^2 = {\rm exp} (i \sqrt{2} \Phi / f) \\
u_{\mu} = i u ^\dagger \partial_{\mu} U u^\dagger  \ .
\end{array}
\end{equation}
The SU(3) matrices for the mesons and the baryons are the
following
\begin{equation}
\Phi =
\left(
\begin{array}{ccc}
\frac{1}{\sqrt{2}} \pi^0 + \frac{1}{\sqrt{6}} \eta & \pi^+ & K^+
\\
\pi^- & - \frac{1}{\sqrt{2}} \pi^0 + \frac{1}{\sqrt{6}} \eta &
K^0 \\
K^- & \bar{K}^0 & - \frac{2}{\sqrt{6}} \eta
\end{array}
\right) \ ,
\end{equation}

\begin{equation}
B =
\left(
\begin{array}{ccc}
\frac{1}{\sqrt{2}} \Sigma^0 + \frac{1}{\sqrt{6}} \Lambda &
\Sigma^+ & p \\
\Sigma^- & - \frac{1}{\sqrt{2}} \Sigma^0 + \frac{1}{\sqrt{6}}
\Lambda & n \\
\Xi^- & \Xi^0 & - \frac{2}{\sqrt{6}} \Lambda
\end{array}
\right) \ .
\end{equation}
In Eq.~(\ref{LagrangN}) $D$ and $F$ are the axial 
vector coupling constants  and $f$ is the 
pseudoscalar meson  decay constant. 
\end{appendix}

\end{document}